\documentclass[%
reprint, superscriptaddress,amsmath,amssymb,aps,
pra,prstper,floatfix]{revtex4-1}

\usepackage[FIGTOPCAP,tight]{subfigure}
\usepackage{graphicx}
\usepackage{dcolumn}
\usepackage{bm}

\usepackage{xcolor}
\usepackage{tensor}
\newcommand{\bra}[1]{\left< #1 \right\vert}
\newcommand{\ket}[1]{\left\vert #1 \right>}
\newcommand{\pare}[1]{\left( #1 \right)}
\newcommand{\abs}[1]{\left\vert #1 \right\vert}
\newcommand{\cor}[1]{\left[ #1 \right]}

\newcommand{\llav}[1]{\left\lbrace #1 \right\rbrace}

\begin{document}

\preprint{APS/123-QED}

\title{Measurement of the temperature of atomic ensembles via \emph{which-way} information}

\author{R. de J. Le\'on-Montiel}
\affiliation{ ICFO-Institut de Ciencies Fotoniques, Mediterranean
Technology Park, Universitat Politecnica de Catalunya, Av. Canal
Olimpic s/n, 08860, Castelldefels, Barcelona, Spain}
\author{Juan P. Torres}
\affiliation{ ICFO-Institut de Ciencies Fotoniques, Mediterranean
Technology Park, Universitat Politecnica de Catalunya, Av. Canal
Olimpic s/n, 08860, Castelldefels, Barcelona, Spain}
\affiliation{Departament of Signal Theory and Communications,
Campus Nord D3, Universitat Politecnica de Catalunya, 08034,
Barcelona, Spain\\}


\begin{abstract}
We unveil the relationship existing between the temperature of an
ensemble of three-level atoms in a $\Lambda$ configuration,
and the width of the emission cone of Stokes photons that are
spontaneously emitted when atoms are excited by an optical
pulse. This relationship, which is based on the amount of
\emph{which-way} information available about where the Stokes
photon originated during the interaction, allows us to put forward
a scheme to determine the temperature of atomic clouds by
measuring the width of the emission cone. Unlike the commonly used
time-of-flight measurements, with this technique, the atomic
cloud is not destroyed during each measurement.
\begin{description}
\item[PACS numbers] 42.50.Ct, 03.67.-a

\end{description}
\end{abstract}
\maketitle

\section{introduction}

Atomic ensembles are key ingredients in many theoretical and
experimental schemes whose aim is the implementation of quantum
information protocols \cite{zoller2001,duan2002}, or the
generation of paired photons with non-classical correlations
\cite{lukin2003,kimble2003}. In these ``writing-reading'' schemes, a weak classical
field (pump pulse) interacts with an atomic ensemble, which leads
to the spontaneous emission of a Stokes photon. Since the Stokes
photon and the atomic ensemble are highly correlated, the
projection of the Stokes photon heralds the generation of an
atomic state that is a coherent superposition of all possible
states of the ensemble where only one atom has been excited, the
so-called collective atomic state \cite{hugues2011}.

The selection of a specific direction of Stokes photons emission
is an important issue, since i) one aims at choosing a direction
that maximizes the flux of generated photons, and ii) the
direction in which the photons are detected might determine
the specific quantum state of the atomic ensemble.

It has been shown that in the case of a room-temperature atomic cloud,
where atoms are considered to move fast within the cloud, Stokes photons
are emitted within a small cone around the direction of propagation of the pump beam
\cite{cirac2002}, whereas in the case where the atoms are
considered to be fixed in their positions ({\em cold atomic
clouds}), Stokes photons have no preferred direction of
emission \cite{scully2003,scully2006}, always that it is not
forbidden by the transition matrix elements. Even though, in most experiments,
the emitted photon is detected at a small angle ($\sim 0^{\circ}- 3^{\circ}$)
\cite{lukin2003,kimble2003,kuzmich2005,kimble2005,inoue2006}.

These results consider only the angular distribution of emitted
photons in two limiting cases: when the atoms are either moving
very fast (high temperature) or completely fixed
(low temperature) within the cloud. Notwithstanding, the transition between these
two cases has not been explored yet. In this paper, we construct a
model to describe the angular distribution of emitted Stokes
photons as a function of the temperature of the atomic cloud. The
importance of our result resides in the fact that we can readily
develop a new technique where the measurement of the width of the
emission cone can be used to determine the temperature of the
atomic cloud. This is made possible by unveiling the close
relationship that exists between the range of possible directions
of emission, and the \emph{which-way} information available about
where the photon originated, i.e., knowledge of the position of
the atom that emitted the Stokes photon during the interaction.

One of the attributes of this new technique is that, unlike
commonly used time-of-flight (TOF) measurements
\cite{lett1988}, the atomic cloud is not destroyed during
each measurement. This new technique is thus added to the group of
non-destructive measurements such as resonance fluorescence
spectrum analysis \cite{lett1990}, recoil-induced resonances
\cite{verkerk1994}, and transient four-wave mixing
\cite{imoto1998}. Also, since this technique does not
require any additional elements in the experimental setup, its
implementation can be easily carried out.

\section{model}

\begin{figure}[t!]
\begin{center}
       {\includegraphics[width=6cm]{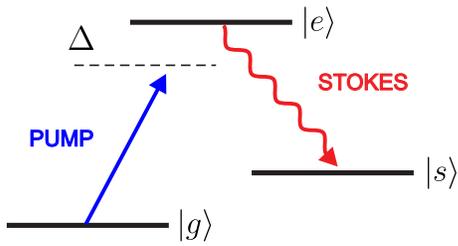}}
\end{center}
\caption{Atomic level configuration.}
\label{figure1}
\end{figure}

Let us consider a cloud of N identical three-level atoms in a
$\Lambda$-configuration (Fig. \ref{figure1}). The cloud is
illuminated by a laser pulse that couples the transition $\ket{g}
\rightarrow \ket{e}$ with a detuning $\Delta$. The spontaneous
decay of the atom ($\ket{e} \rightarrow \ket{s}$) leads to the
generation of a photon with different wavelength (Stokes photon).
Here we investigate two important features that, as we will see
later, depend on the temperature of the cloud. First, the angular
distribution of the spontaneously emitted photons, and second, the heralded
generation of the symmetric collective atomic state after detection
of the Stokes photon.

The pump beam, which is a slowly-varying classical field
propagating along the $z$-direction, with a Rayleigh range much
larger than the length of the atomic cloud, writes
\begin{equation}
E_{p}\left({\bf r},t\right)= u\left({\bf r}_{\perp} \right) \xi(t)
\exp\left\{ i k_{0}z - i\omega_{0}t\right\} + h.c. ,
\end{equation}
where $\omega_{0}=k_0 c$ is the central frequency, $c$ is velocity
of light, $u\left( {\bf r}_{\perp}\right)$ describes the
transverse spatial shape of the pump beam and $\xi(t)$ its temporal
shape.

The Stokes field, which is described quantum mechanically, writes
\begin{equation}
\hat{E}_{s}^{\dagger}\left( {\bf r},t\right) = \int \hat{a}({\bf
k}) \exp\left\{i {\bf k} \cdot {\bf r} - i\omega t \right\} d{\bf
k},
\end{equation}
where $\hat{a}({\bf k})$ is the annihilation operator, ${\bf k}=(k_{x},k_{y},k_{z})$ is the wavevector
of the Stokes photon and $\omega=|{\bf k}| c$ its frequency.

In the interaction picture, and after adiabatically eliminating
the upper level $\ket{e}$, the interaction of the light field with
the atomic cloud can be described by the Hamiltonian
\cite{cirac2002}
\begin{equation}
\label{hamiltonian}
\begin{split}
H(t)=&\sum_{i=1}^{N} \hat{\sigma}_{sg}^{i} \int d{\bf k} g_{\bf k}
a^{\dagger}({\bf k}) \exp\left\{ i\Delta\omega t \right\}
u\left({\bf r}_{\perp,i}\right)\xi\pare{t} \\
& \times \exp\left\{ -i\Delta {\bf k} \cdot {\bf r}_{i}\right\} +
h.c.,
\end{split}
\end{equation}
where $\hat{\sigma}_{sg}^{i}=\ket{s}_{i}\bra{g}$ is the transition
operator for the \textit{i}th atom, ${\bf r}_{i}=(x_{i},y_{i},z_{i})$ is the vector
position of the \textit{i}th atom, $g_{{\bf k}}$ is the coupling
coefficient of the transition, $\Delta\omega = \omega -
\pare{\omega_{0} - \omega_{sg}}$ and $\Delta {\bf k} = {\bf k} -
k_{0}\hat{z}$, with $\omega_{sg}$ being the transition frequency
between states $\ket{g}$ and $\ket{s}$.

Before the interaction, we consider that all the atoms are in the ground state and
that there are no photons in the optical modes, i.e,
$\ket{\Psi}_{0}= \ket{g_{1}...g_{i}...g_{N}}\otimes\ket{0}_{\bf k}$.
Then, considering that the pump field is weak enough, we can make
use of first-order perturbation theory to write the state of the
system as
\begin{equation}
\label{state_atoms2}
\begin{split}
\ket{\Psi} =&\ket{\Psi}_{0}-i\;\varepsilon\left(\Delta \omega
\right)\sum_{i=1}^{N}\int d{\bf k}\; u\left(
{\bf r}_{\perp,i}\right)  \\
& \times \exp\left\{-i\Delta {\bf k}\cdot {\bf
r}_{i}\right\}\ket{g_{1}...s_{i}...g_{N}}\vert{\bf k}\rangle,
\end{split}
\end{equation}
where $\varepsilon(\Delta \omega)=\int_{0}^{t} dt' g\hspace{0.8mm}\xi(t')
\exp(i\Delta \omega t')$. We assume that $\Delta {\bf k}$ becomes independent of the
frequency and that $g=g_{{\bf k}}$ has
the same value for all allowed directions of emission of the
Stokes photons.

The use of perturbation theory is motivated by
experiments in which a weak pump pulse and a short interaction time are used in order to guarantee that the
probability of creating more than one excitation in the collective atomic state is very low \cite{kimble2003,kuzmich2005,kimble2005,inoue2006}. This weak-pumping condition makes a perturbative approach suitable for describing a realistic situation.

In the case of cold atomic ensembles, since the atoms are
considered to be fixed in their positions, we can directly use Eq.
(\ref{state_atoms2}) to obtain that the probability of emitting a
photon in a given direction ${\bf k}$ is equal for all directions,
i.e., there is no preferred direction of emission
\cite{scully2003,scully2006}, independently of the specific shape
of the atomic cloud, always that it is not forbidden by the
transition matrix elements.

In the case of hot atomic ensembles, due to the fact that during
the light-atom interaction time the atoms are moving fast, an
average value over all positions ${\bf r}_{i}$ should be performed
\cite{cirac2002,eberly2006}. The resulting state becomes
\begin{equation}
\label{state_hot_atoms}
\begin{split}
\ket{\Psi}= & \ket{\Psi}_{0}-i\varepsilon\left(\Delta \omega
\right)\int d{\bf k}\; F\left(\Delta {\bf k}\right) \vert{\bf
k}\rangle \otimes \sum_{i=1}^{N} \ket{g_{1}...s_{i}...g_{N}}
\end{split}
\end{equation}
where the average value over all positions is expressed as
$F\left(\Delta {\bf k}\right)= \int d{\bf r} u({\bf
r}_{\perp})\exp \left\{-i\Delta {\bf k}\cdot {\bf r}\right\}
p_{dis}({\bf r})$, with $p_{dis}({\bf r})$ being the atomic
distribution function. From Eq. (\ref{state_hot_atoms}), one can
show that the Stokes photons are emitted in a small cone around
the forward direction (see \cite{cirac2002} for a detailed
calculation), whose width depends on the particular spatial shape
of the atomic cloud. Notice that, in this case, the photon
and atomic degrees of freedom can be decoupled, and the quantum
state of the atoms is the so-called symmetric collective atomic state,
i.e., $\ket{s_a}=1/\sqrt{N}\,\sum_{1}^{N}
\ket{g_{1}...s_{i}...g_{N}}$.

Equation (\ref{state_atoms2}) does not show any temperature
dependence, so in its present form cannot be used to describe the
transition between the two limiting scenarios considered so far:
warm and cold atomic ensembles. In order to model the temperature
dependence, we introduce a function that describes the movement of
each atom around its mean position ${\bf r}_{i}$,
\begin{equation}
\label{function_new} f\pare{{\bf r},{\bf r}_{i}} =
\frac{1}{\pi^{3/2}A^{3}\pare{T}} \exp \cor{-\frac{\abs{{\bf r} -
{\bf r}_{i}}^{2}}{A^{2}\pare{T}}},
\end{equation}
where $A\pare{T} = v_{a}\tau$ determines the radius of the area
over which the atoms can move during the interaction time. It
depends on the pump pulse duration ($\tau$), and on the speed
($v_{a} = \sqrt{2K_{B}T/m}$) most likely to be possessed by any
atom of the system. Here $m$ is the mass of the atom, $K_{B}$ is
the Boltzmann constant and $T$ is the temperature of the atomic
ensemble. Notice that the origin of $v_{a}$ lies in the
Maxwell$-$Boltzmann distribution. This distribution is assumed, because
it has been shown that the Maxwell-Boltzmann distribution provides an accurate description of the motion of atoms at temperatures above tenths of $\mu K$ \cite{lett1988,lett1989}. Hence, Eq. (6) is useful for describing the motion of atoms undergoing a transition from the hot to the cold condition, provided that the lowest temperature values are above tenths of $\mu K$.

Making use of the function given in Eq. (\ref{function_new}) to
rewrite Eq. (\ref{state_atoms2}), the temperature-dependent
quantum state of the system atoms-photon can be written as
\begin{equation}
\begin{split}
\label{new_state} \ket{\Psi} = &
\ket{\Psi}_{0}-i\varepsilon\left(\Delta \omega
\right)\sum_{i=1}^{N}\int d{\bf k} \int_{V} d{\bf r}
f\pare{{\bf r},{\bf r}_{i}} u\left( {\bf r}_{\perp}  \right)\\
&  \times  \exp\left\{ -i\Delta {\bf k}\cdot {\bf r} \right\}
\ket{g_{1}...s_{i}...g_{N}} \vert{\bf k}\rangle,
\end{split}
\end{equation}
where $V$ is the volume of the cloud.

Note that, in the limit where $A\rightarrow 0$, the function given
in Eq. (\ref{function_new}) tends to a Dirac delta function, and
we recover the state of the system described by Eq.
(\ref{state_atoms2}).

In order to obtain the angular distribution of the emitted Stokes
photons, we trace out the atomic variables of the density matrix
of the system ($\rho = \ket{\Psi}\bra{\Psi}$). Neglecting the vacuum
contribution, the reduced density matrix of the photon state writes
\begin{equation}
\label{density_matrix_stokes} \rho_{s} =  \sum_{i=1}^{N} \int
d{\bf k}d{\bf k}'S({\bf r}_{i},{\bf k})S^{*}({\bf r}_{i},{\bf k'})
\vert{\bf k}\rangle \langle {\bf k}'\vert ,
\end{equation}
where
\begin{equation}
S({\bf r}_{i},{\bf k}) = \int_{V} d{\bf r} f({\bf r},{\bf r}_{i})
u({\bf r}_{\perp})\exp\left(-i\Delta {\bf k}\cdot {\bf r} \right).
\end{equation}

Considering that the atoms are contained in a
cell with transversal dimensions $L_{x}$, $L_{y}$ and
length $L_{z}$, we can solve Eq. (9) to obtain
\begin{widetext}
\begin{equation}
S({\bf r}_{i},{\bf k}) = \frac{1}{8}\alpha^{2}\Phi(x_{i},k_{x})\Phi(y_{i},k_{y})\Omega(z_{i},k_{z}),
\end{equation}
where
\begin{eqnarray}
\alpha&=& \pare{\frac{r_{0}^{2}}{A^{2} + r_{0}^{2}}}^{1/2},  \\
\Phi(x_{i},k_{x}) &=& \exp\cor{-\frac{1}{4}k_{x}^{2}r_{0}^{2} - \cor{\frac{\alpha}{r_{0}}\pare{x_{i} + i\frac{1}{2}r_{0}^{2}k_{x}}}^{2}} \nonumber \\
& & \times \llav{\mathrm{erf}\cor{-\frac{1}{2}\alpha^{3}\pare{2x_{i}-ik_{x}A^{2}-\frac{L_{x}}{\alpha^{2}}}} - \mathrm{erf}\cor{-\frac{1}{2}\alpha^{3}\pare{2x_{i}-ik_{x}A^{2}+\frac{L_{x}}{\alpha^{2}}}}}, \\
\Omega(z_{i},k_{z}) &=& \exp\cor{-\frac{1}{4}k_{z}^{2}A^{2} - ik_{z}z_{j}} \nonumber \\
& & \times \llav{\mathrm{erf}\cor{-\frac{1}{2A}\pare{2z_{i}-ik_{z}A^{2}-L_{z}}}-\mathrm{erf}\cor{-\frac{1}{2A}\pare{2z_{i}-ik_{z}A^{2}+L_{z}}}}.
\end{eqnarray}
\end{widetext}

Notice that the presence of the error function (erf(x)) in Eq. (10) is due to the
integration over the finite volume (V) of the cell that contains the atoms.

We now find that the probability of emitting a Stokes photon in the direction ${\bf
k}$ is given by the diagonal terms of the density matrix
(\ref{density_matrix_stokes}),
\begin{equation}
\label{probability_of_emission} P({\bf k}) = \sum_{i=1}^{N}
\abs{S\left( {\bf r}_{i},{\bf k}\right)}^2 ,
\end{equation}
where the normalization condition writes $\sum_{i=1}^{N} \int
d{\bf k}\abs{S\left({\bf r}_{i},{\bf k}\right)}^2=1$. In general,
since the atomic cloud contains a large atom number density, the
atomic summation can be rewritten as $\sum_{i=1}^{N} \rightarrow
(N/V)\int dV$.

Note that Eq. (14) needs to be solved numerically due to the
presence of the error function in Eq. (10). Notwithstanding, since the functions of the
spatial variables are separated (as can be seen in Eq. (10)), the numerical integration of Eq. (\ref{probability_of_emission})
can be easily performed.

\section{angular distribution of emitted Stokes photons}

We have calculated the angular distribution of the emitted Stokes
photons considering an ensemble of $^{87}$Rb atoms contained in a
pencil-shaped cell with transversal dimensions: $L_{x}=L_{y}=2$
mm, and length $L_{z}=30$ mm. The atoms are illuminated by a pump
pulse with a transversal shape given by $u({\bf r}_{\perp}) \sim
\exp\left\{ -(x^2 + y^2)/r_{0}^2\right\}$, where $r_{0} = 2$ mm is
the beam waist of the pump beam. The level configuration of the
atoms is set to $5^{2}P_{1/2}$ for the excited level $\ket{e}$,
and the Zeeman-splitting levels $5^{2}S_{1/2} \pare{F=1}$ and
$5^{2}S_{1/2} \pare{F=2}$ for the $\ket{g}$ and $\ket{s}$ states,
respectively.

\begin{figure}[!t]
    \begin{center}
       \includegraphics[width=7cm]{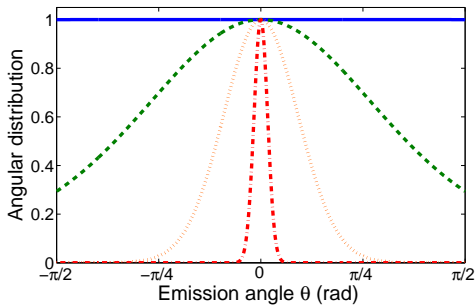}
    \end{center}
\caption{Angular distribution of emitted Stokes photons for
different temperatures of the atomic cloud. Solid line: $T=100$
$\mu$K; Dashed line: $T=1$ K; Dotted line: $T=10$ K and
Dash-dotted line: $T=300$ K. In all cases, the pump pulse duration
is set to $\tau=10$ ns.}
\end{figure}

Figure 2 shows the angular distribution (normalized to the
maximum) of emitted Stokes photons as a function of the angle
($\theta$) between the direction of the pump and the emitted
photon (as shown in Fig. 3(a)). In the low temperature limit, the
spontaneous emission of Stokes photons has no preferred direction. This result agrees with
ref. \cite{kimble2003}, in which Stokes photons are said to be emitted into
$4\pi$ steradian. In contrast, as the temperature of the cloud is increased, the
probability distribution narrows around $\theta=0^{\circ}$,
evidencing the fact that, in the case of warm atomic ensembles,
Stokes photons are emitted preferentially along the direction of
the pump, as it has been experimentally observed, for instance in
\cite{lukin2003}.

The results shown in Fig. 2 can be understood in terms of the which-way information left in the atoms after emitting a Stokes photon, i.e., information about the position of the atom that emitted the photon. On one hand, for the case of cold atoms, due to the fact that they are fixed, one can obtain, in principle, information about the position of the atom that emitted the photon. In this situation, the possible paths of the Stokes photon will not interfere, because which-way information has been left in the atomic ensemble. This can be clearly seen from Eq. (14), which for cold atomic ensembles takes the form
\begin{equation}
P_{cold}({\bf k}) = \sum_{i=1}^{N} \abs{u\left({\bf r}_{\perp,i}\right)}^{2}.
\end{equation}

Equation (15) shows that emission of Stokes photons from a cold atomic ensemble has no preferred direction.

On the other hand, notice that for the case of hot atomic ensembles, Eq. (6) is a constant inside the integration volume, so we can write Eq. (14), by means of the large atom number density relation, as
\begin{equation}
P_{hot}({\bf k}) = \abs{\sum_{i=1}^{N} u\left({\bf r}_{\perp,i}\right)e^{-i\Delta {\bf k}\cdot {\bf
r}_{i}}}^{2}.
\end{equation}

We see from Eq. (16) that interference between the possible paths of the Stokes photon is now restored, because which-way information has been erased by the movement of the atoms in the cloud. Interestingly, this which-way information effect has been also observed, for instance, in the context of second-order interference of single photons \cite{mandel1991}.

\begin{figure}[!t]
    \begin{center}
       \subfigure[]{\hspace{6mm}\includegraphics[width=7cm]{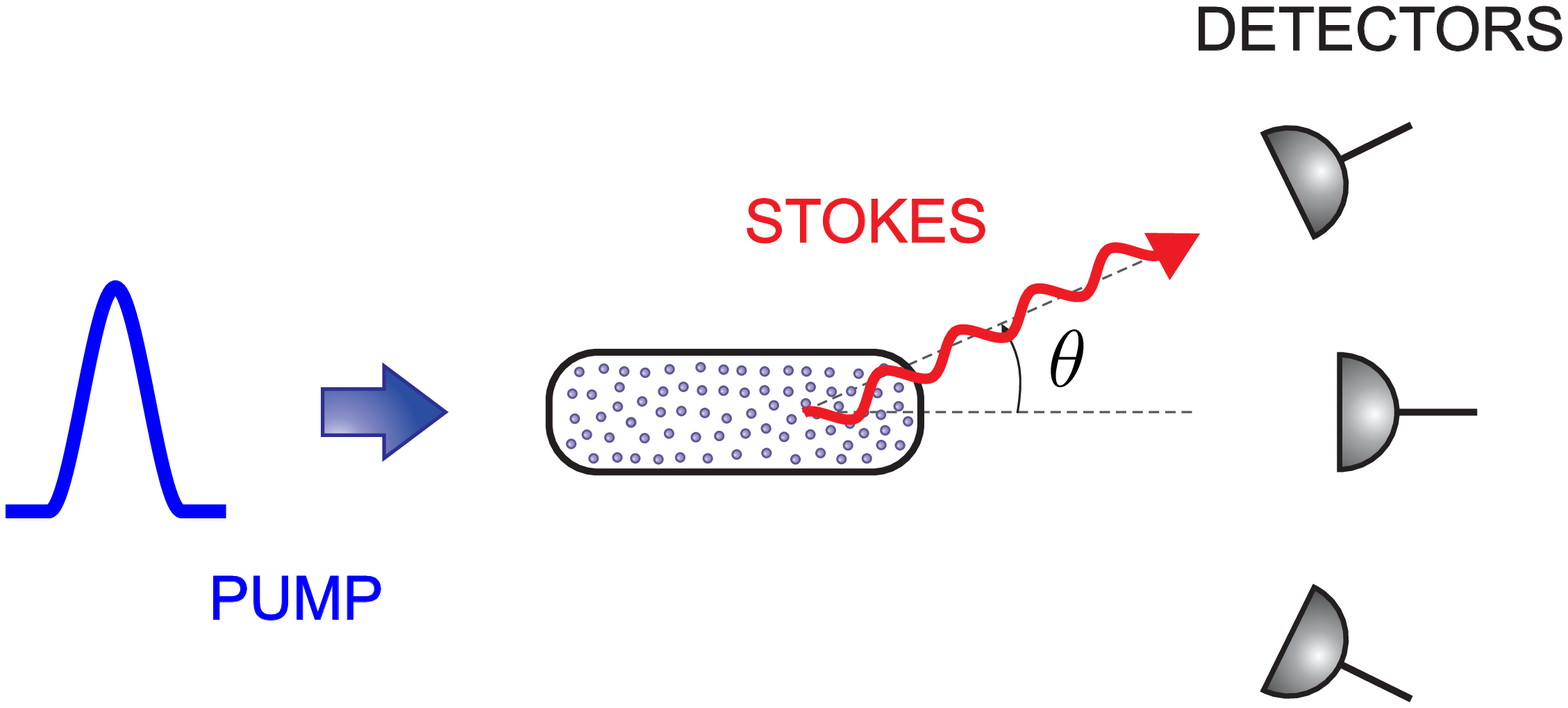}}  \\ \vspace{0mm}
       \subfigure[]{\includegraphics[width=6.25cm]{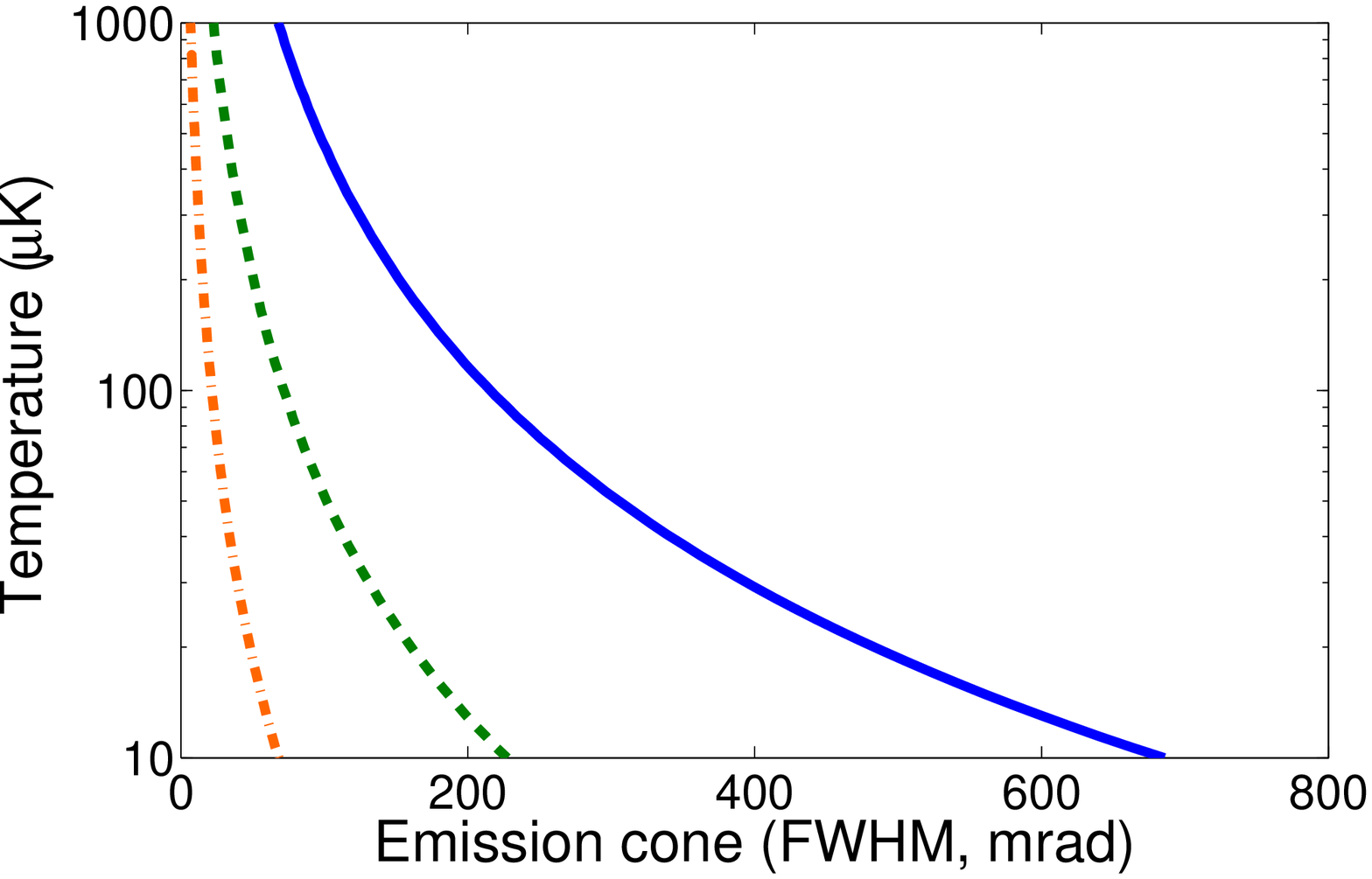}}    
    \end{center}
\caption{(a) Proposed experimental setup: an array of detectors
is used to measure the width of Stokes emission cone, which
allows to determine the temperature of the atomic ensemble. (b)
Temperature of the atomic cloud as a function of the FWHM of
the emission cone for different pulse durations. Solid line:
$\tau=10$ $\mu$s; Dashed line: $\tau=30$ $\mu$s; Dash-dotted line:
$\tau=100$ $\mu$s. }
\end{figure}

It is important to highlight that, it is not the actual acquisition of information from the system which determines the rise of interference, but that in principle, it would be possible to obtain such information \cite{mandel1999}.

\section{experimental proposal}

The close relationship between the width of the emission cone and
the temperature of the atomic ensemble allows us to put forward a
new technique to determine the temperature of atomic clouds. The
proposed experimental setup consists of an array of detectors (or
a movable detector) that would be able to detect Stokes photons along
different directions, as depicted in Fig. 3(a). In this way, by
measuring the width of the emission cone, we can make use of Eq.
(\ref{probability_of_emission}) to retrieve information about
the temperature of the atomic ensemble. Figure 3(b) shows the
temperature of the atomic ensemble as a function of the full width
at half maximum (FWHM) of the emission cone. Notice that, by
choosing a sufficiently short pulse, the relationship between the
emission cone width and the temperature gets smoother. This can be
useful for a better discrimination of the width of the emission
cone, enhancing thus the precision of the technique. Also, notice that this
technique is not based on the ballistic expansion of the atomic cloud \cite{lett1988}, so each measurement can be
performed without destroying it.

\section{Heralded generation of the symmetric atomic state}

We can also use Eq. (7) to describe how the generation of the symmetric state depends on
the temperature of the atomic cloud. When the Stokes photon is detected in an arbitrary
direction ${\bf k}$, i.e., is projected into the state $\vert {\bf k}\rangle$, the
corresponding quantum state of the atomic cloud is
\begin{equation}
\ket{\Psi}_{a} = \sum_{i=1}^{N} S\left( {\bf r}_{i},{\bf k}\right)
\ket{g_{1}...s_{i}...g_{N}}.
\end{equation}

\begin{figure}[!t]
    \centering
        \subfigure[]{\raisebox{0mm}{\includegraphics[width=4cm]{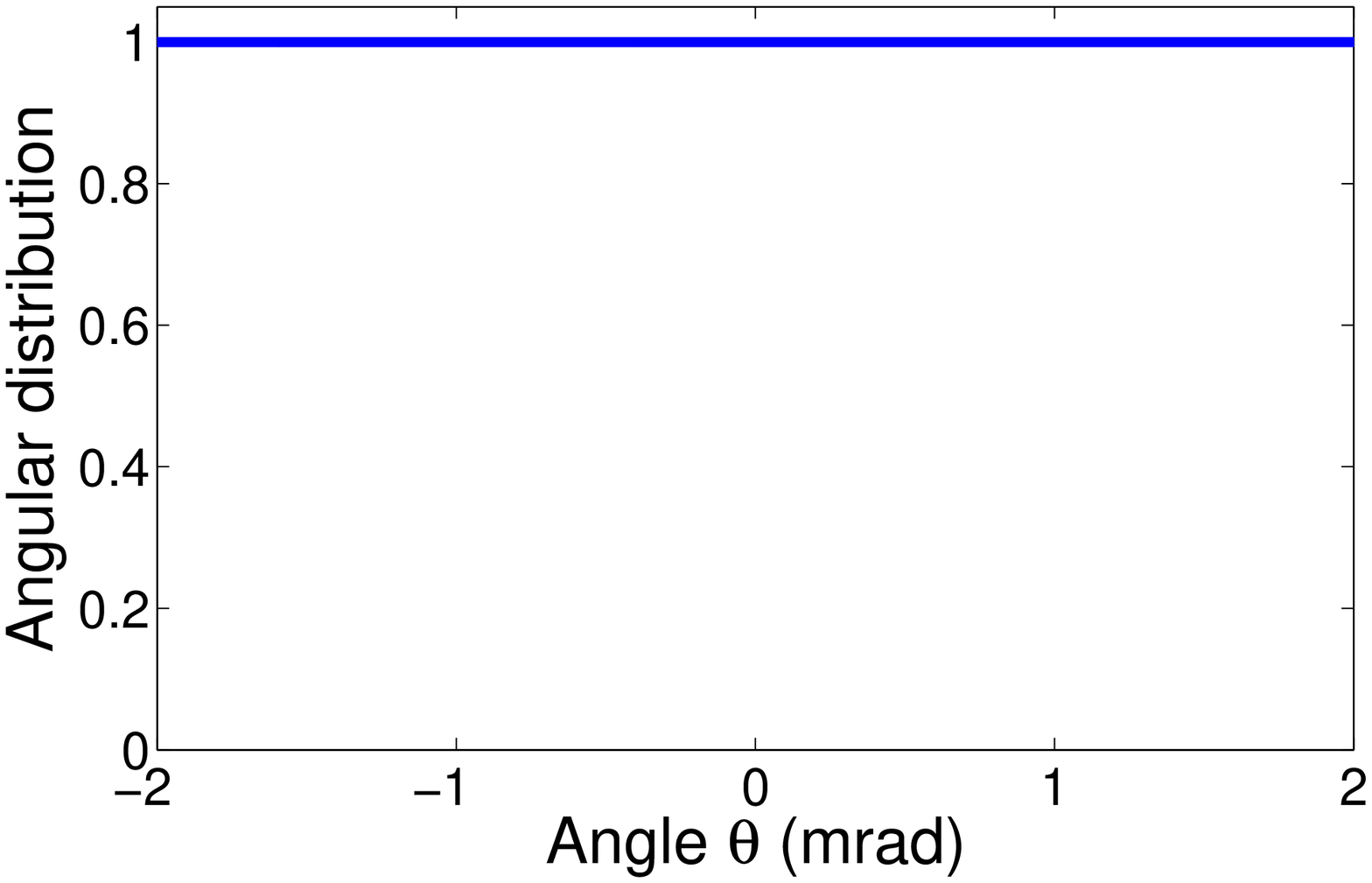}}}
        \subfigure[]{\raisebox{0mm}{\includegraphics[width=4cm]{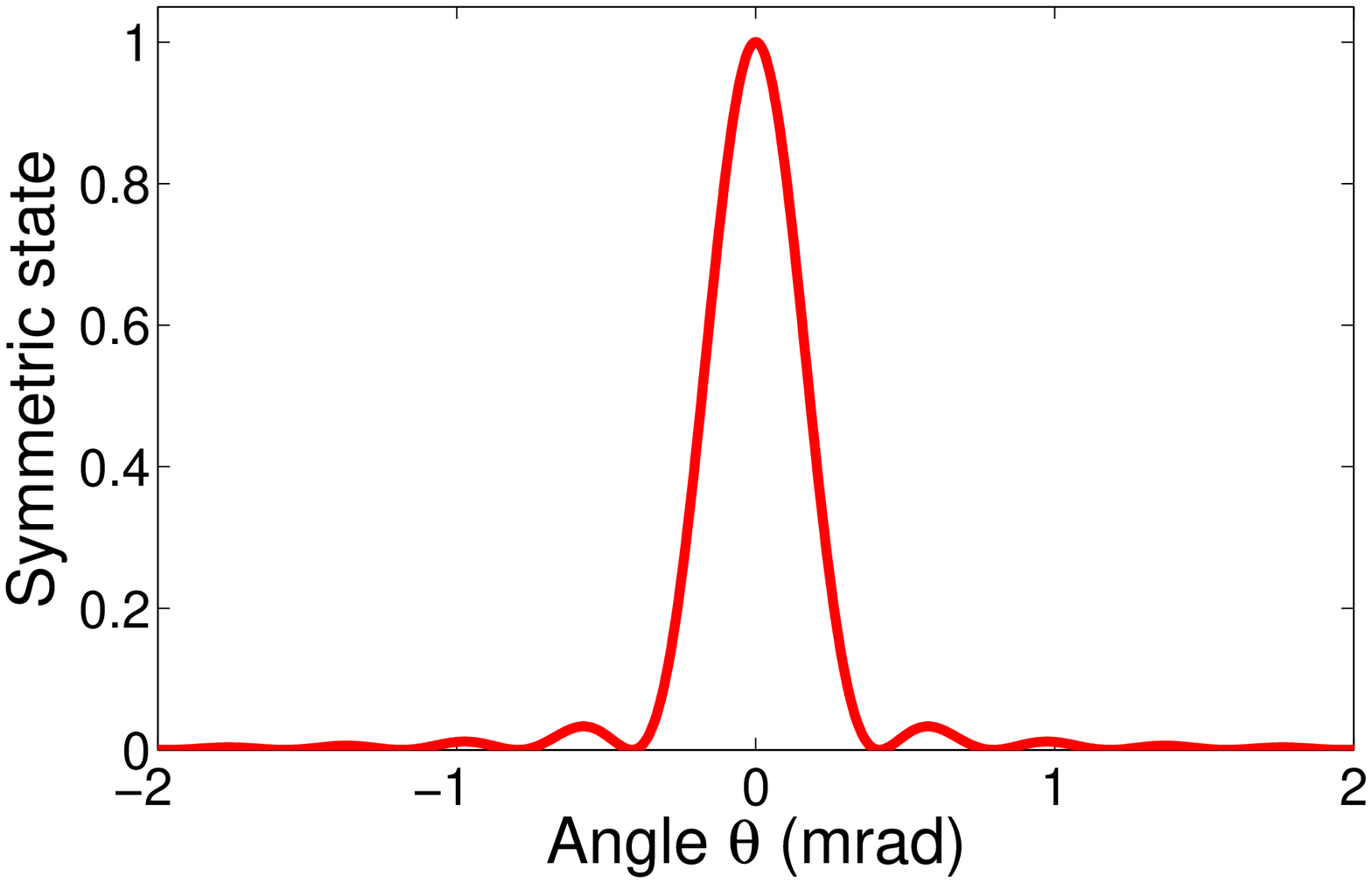}}}\\
        \subfigure[]{\raisebox{0mm}{\includegraphics[width=4cm]{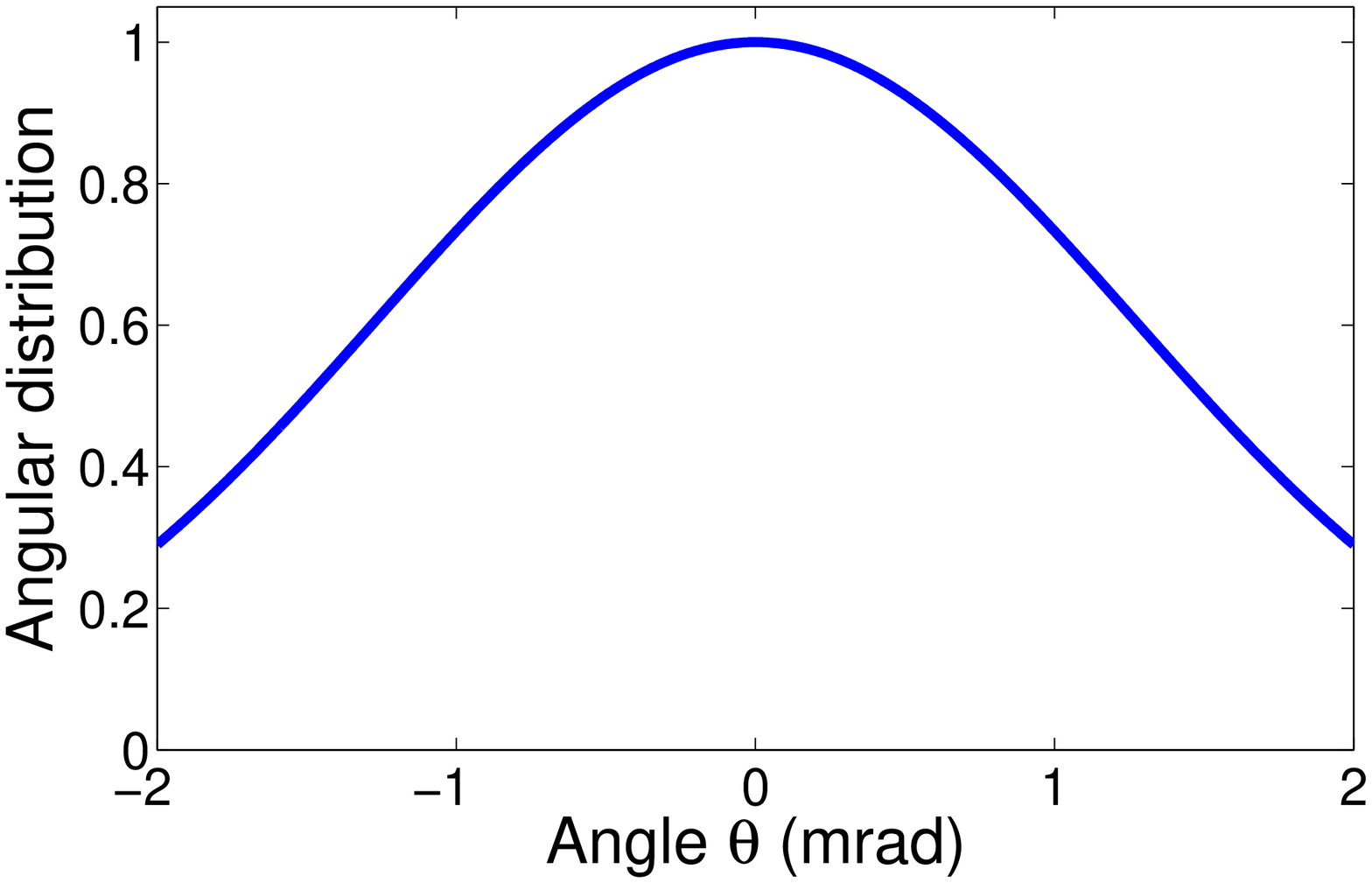}}}
        \subfigure[]{\raisebox{0mm}{\includegraphics[width=4cm]{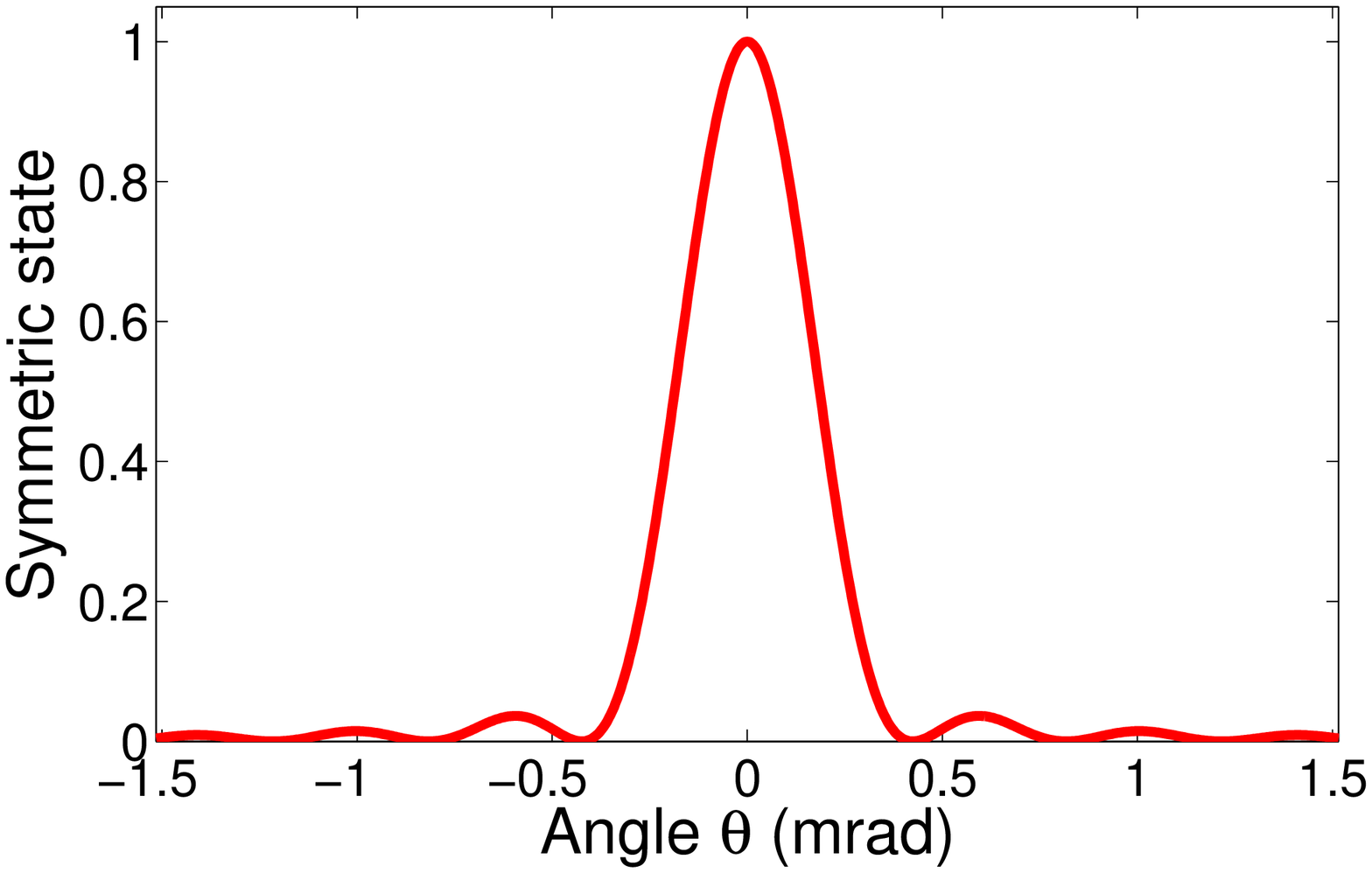}}}\\
        \subfigure[]{\raisebox{0mm}{\includegraphics[width=4cm]{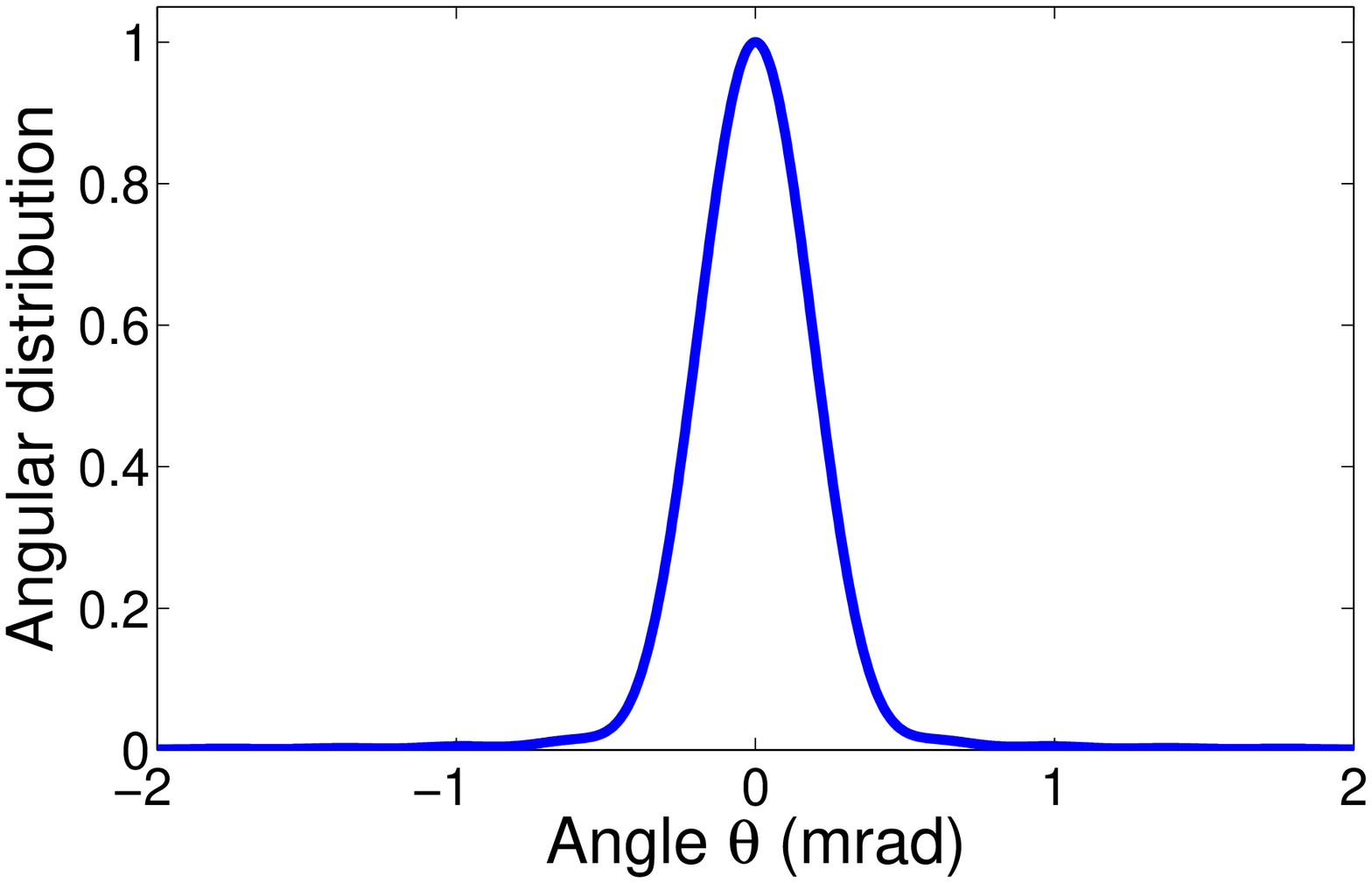}}}
        \subfigure[]{\raisebox{0mm}{\includegraphics[width=4cm]{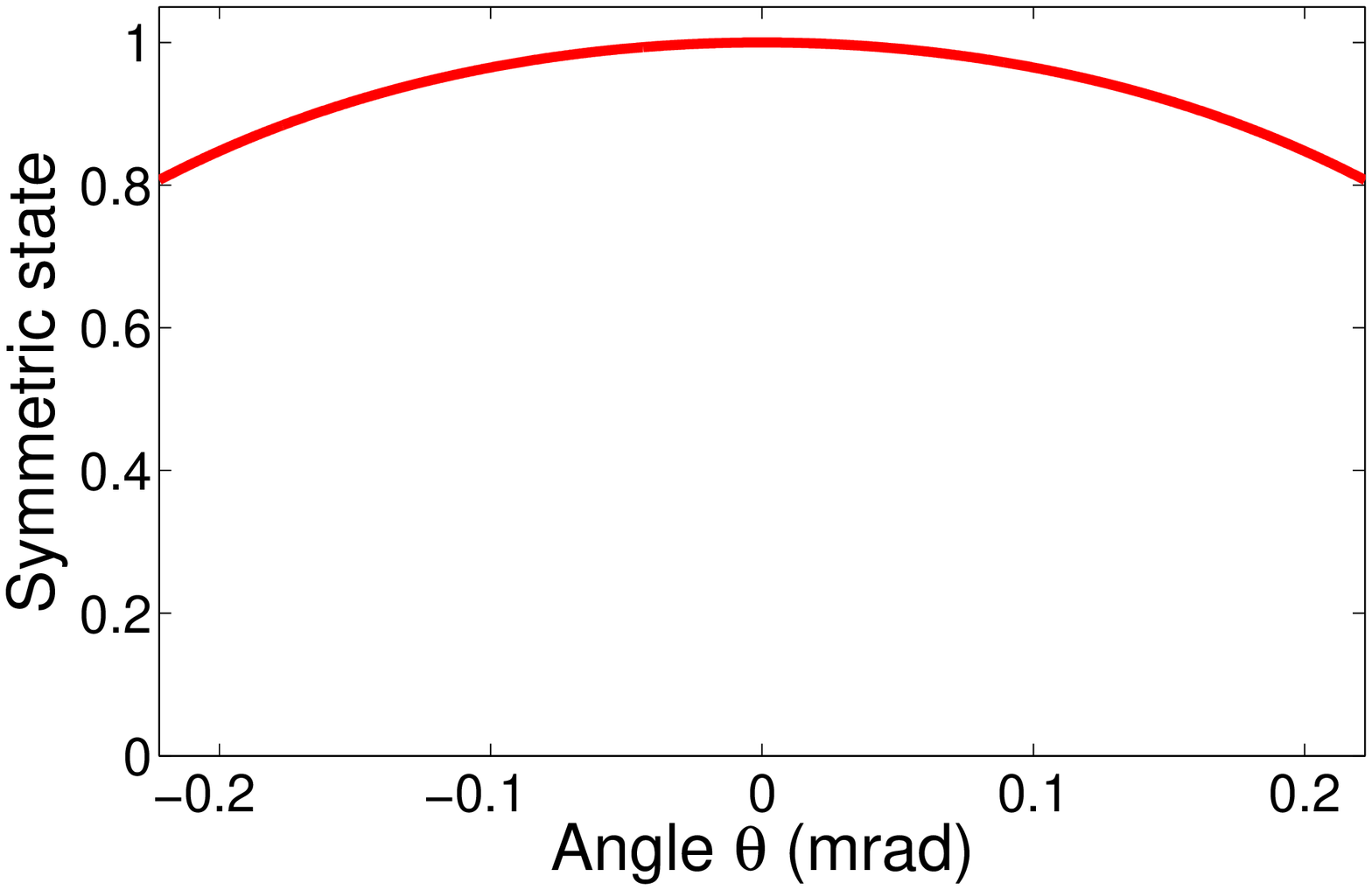}}}\\
        \subfigure[]{\raisebox{0mm}{\includegraphics[width=4cm]{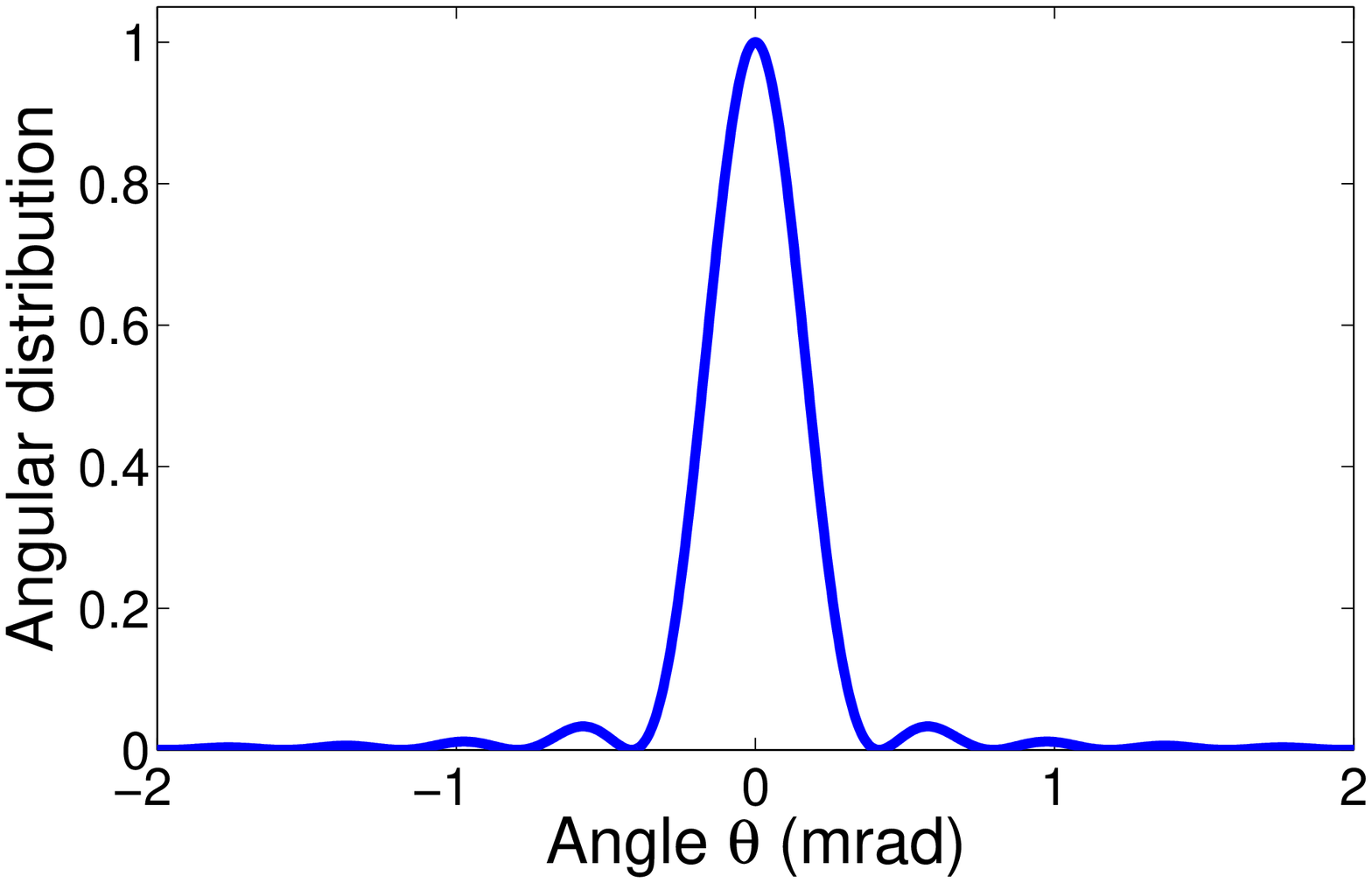}}}
        \subfigure[]{\raisebox{0mm}{\includegraphics[width=4cm]{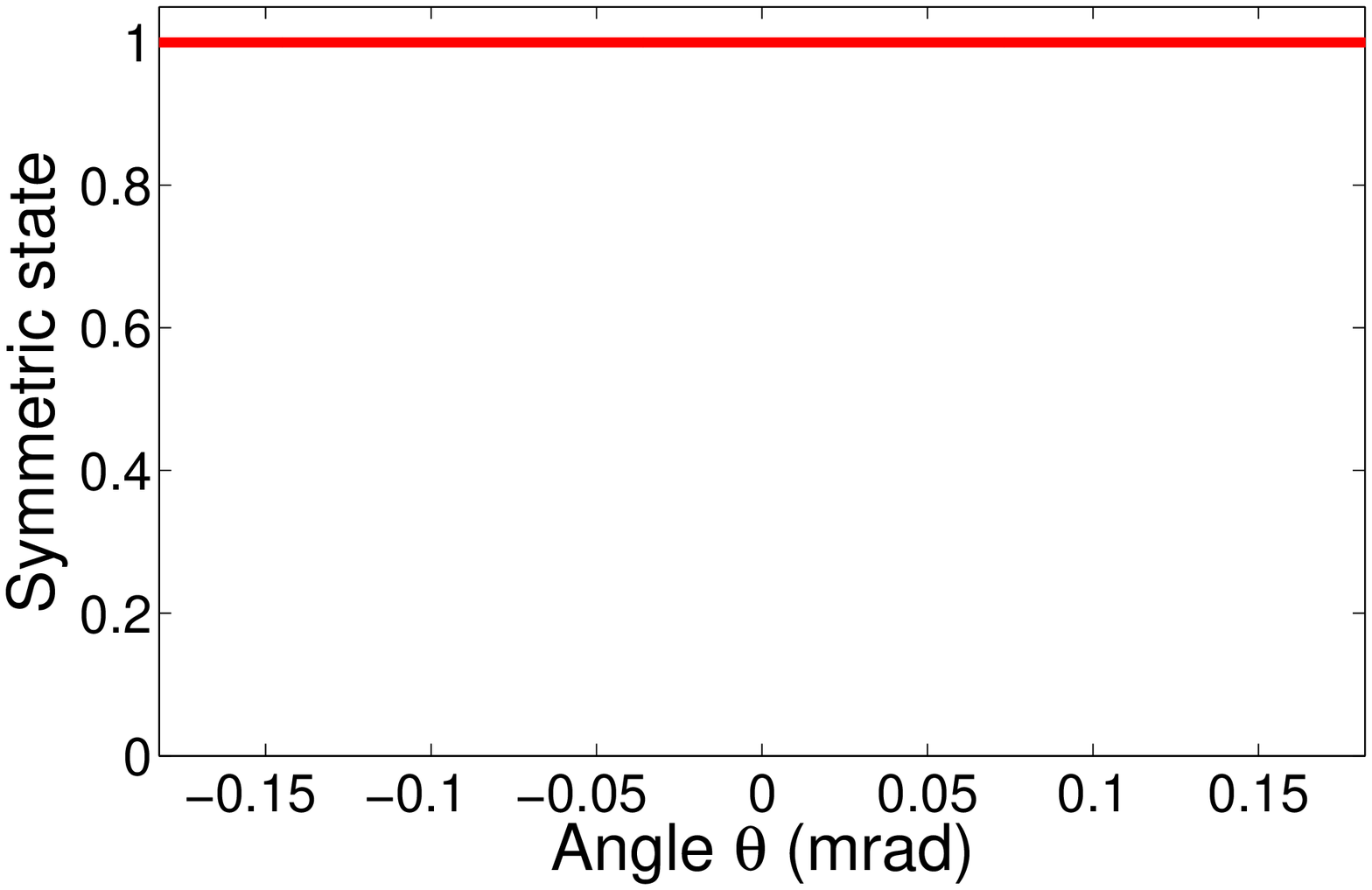}}} \\
        \caption{Angular distribution of the emitted Stokes photons ((a),(c),(e) and (g)),
        and the weight of the symmetric atomic state $\abs{\left\langle s_a\vert\Psi \right\rangle_{a}}^2$ (within the FWHM
        of the emission cone) ((b), (d), (f) and (h)) for different radius
        of the region where the atoms are let to move: (a,b) A=1 $\mu$m,
        (c,d) A=100 $\mu$m, (e,f) A=1 mm, and (g,h) A=100 mm. Pulse duration: 10 $\mu$s.}
\label{figure4}
\end{figure}

In the case where the atoms are barely moving during the
interaction time, only a small fraction of generated photons
(those in a small angle around the pump beam direction)
corresponds to the symmetric state, even though photons are
emitted in a larger emission cone. This can clearly be seen by
comparing Figs. \ref{figure4}(a) and (b), where we plot the
projection $\abs{\left\langle s_a\vert\Psi \right\rangle_{a}}^2$.
When the atoms are let to move within the cloud by increasing the
temperature, the \emph{which-way} information is erased and the
emission cone gets narrower (see Fig. \ref{figure4}(g)). In this
case, as it can be seen from Fig. \ref{figure4}(h), photons
emitted in all allowed possible directions are in the symmetric
state. Notwithstanding, we are again forced to consider small
emission angles around the pump beam direction to enhance the flux
of detected Stokes photons. Therefore, in all cases, photons should
be collected in a small cone around the direction of propagation
of the pump beam, if the goal is to generate the symmetric atomic
state. But as Fig. \ref{figure4} shows, the reason behind this
restriction depends on the temperature of the atomic ensemble.

\section{conclusion}

In this paper, we have presented a new technique to measure the
temperature of atomic ensembles, based on the relationship between
the temperature of the ensemble and the width of the emission cone
of the spontaneously emitted Stokes photons. We have also shown that
heralded generation of the collective symmetric atomic state
requires the detection of the heralding Stokes photon in a narrow
cone around the direction of the exciting pulse.
For cold atomic ensembles, this is the only direction of emission
that guarantees the generation of such state, while for warm
ensembles, it is the direction with the highest
efficiency.

\acknowledgments

This work was supported by projects FIS2010-14831 and FET-Open 255914
(PHORBITECH). This work has also been supported by Fundacio Privada Cellex Barcelona.
We thank J. Svozil\'{i}k and G. Puentes for helpful discussions.

\end{document}